\documentclass[%
 reprint,
superscriptaddress,
 amsmath,
 amssymb,
 aps,
 prd,
]{revtex4-2}

\usepackage{amsmath}
\usepackage{amssymb}
\usepackage{graphicx}
\usepackage[caption=false]{subfig}
\usepackage{enumitem}
\usepackage{color}
\usepackage[percent]{overpic}
\usepackage{bm}
\usepackage{rotating}
\usepackage{cancel}
\usepackage{comment}
\usepackage{braket}
\usepackage{wasysym}
\usepackage{mathtools}
\usepackage{hyperref}
\usepackage{tikz}
\usetikzlibrary{snakes}
\usetikzlibrary{external}

\usepackage{dcolumn}% Align table columns on decimal point
\usepackage{amsthm} % Theorem Formatting
\usepackage{simplewick}

\allowdisplaybreaks

\newcommand{\ketbra}[2]{\left| #1 \vphantom{#2}\right>\!\!\left< #2\vphantom{#1}\right|}
\newcommand{\abs}[1]{\left| #1 \right|} % for absolute value

\newcommand{\ii}{\mathrm{i}}

\newcommand{\dd}{\mathrm{d}}
\newcommand{\px}{+}
\newcommand{\mx}{-}

\newcommand{\conjnf}[1]{#1^{*}}
\newcommand{\Tr}[0]{\text{Tr}}

\begin{document}

\title{Gaussian states in quantum field theory: Exact representations of relative phase in superpositions of Gaussian states}

\author{Nicholas Funai}
 %\altaffiliation[Also at ]{Physics Department, XYZ University.}%Lines break automatically or can be forced with \\
%\author{Second Author}%
% \email{Second.Author@institution.edu}
\affiliation{Centre for Quantum Computation and Communication Technology, School of Science,
RMIT University, Melbourne, Victoria 3001, Australia}%
%\affiliation{Institute for Quantum Computing, University of Waterloo, Waterloo, Ontario, N2L 3G1, Canada}
%\affiliation{Department of Applied Mathematics, University of Waterloo, Waterloo, Ontario, N2L 3G1, Canada}
%
%\author{Nicolas Menicucci}
%\author{Ben Baragolia}
%\affiliation{Institute for Quantum Computing, University of Waterloo, Waterloo, Ontario, N2L 3G1, Canada}
%\affiliation{Department of Applied Mathematics, University of Waterloo, Waterloo, Ontario, N2L 3G1, Canada}
%\affiliation{Perimeter Institute for Theoretical Physics, 31 Caroline St N, Waterloo, Ontario, N2L 2Y5, Canada}

%\date{\today}

\begin{abstract}
Gaussian quantum mechanics is a powerful tool regularly used in quantum optics to model linear and quadratic Hamiltonians efficiently. Recent interest in qubit-CV hybrid models has revealed a simple, yet important gap in our knowledge, namely, how to fully manipulate superpositions of Gaussian states. In this paper, we show how to faithfully represent a quadratic Gaussian state in the Fock basis. Specifically, we evaluate the phase necessary to equate the unitary representation of a zero-mean Gaussian state with its Fock representation. This allows for the coherent manipulation of superpositions of Gaussian states, especially the evaluation of expectation values from these superposed states. We then use this method to model a simple quantum field theory communication protocol using quadratic detectors in a regime that has previously been impossible to solve. The result presented in this paper is expected to become increasingly relevant with hybrid-CV systems, especially in the strong-coupling regime.
\end{abstract}

\maketitle

%%%%%%%%%%%%%%%%%%%%%%%%%%%%%%%%%%%%%%%%%%%%%
\section{Introduction}
Quantum optics (QO) is a curious corner of quantum physics where inherently relativistic electromagnetic fields are simplified under a host of reasonable approximations~\cite{Scully_Zubairy_1997}, resulting in optical modes that are modelled as first quantised harmonic oscillators. The identification of individual optical modes as harmonic oscillators and the predominance of quadratic interactions has made quantum optics synonymous with Gaussian quantum mechanics (QM)~\cite{Gaussian_QM_review,e_farming,brask2022gaussianstatesoperations}. Within the realm of optics and other photon based experiments this has previously been considered as sufficient, although experimental proposals (e.g. optomechanics~\cite{optomech_1,opto_cavendish,optomech_3}) and quantum information demands (e.g. Gottesman-Kitaev-Preskill (GKP) states~\cite{GKP_orig,GKP_cQED}) are pushing beyond the domain of Gaussian quantum mechanics. 

The power and versatility of Gaussian quantum mechanics has applications beyond quantum optics, e.g. in quantum field theory (QFT)~\cite{e_farming,Polo-Gomez_delta_qubit}. However, as experimental techniques and control improve, the assumptions underlying QO and Gaussian QM are becoming less relevant, and an extended theoretical basis for near-Gaussian quantum theory will soon be required~\cite{cQED_harvest}. A previously studied example involves the use of a parametric down converter (PDC)~\cite{squeeze_pdc_0}, usually pumped with a coherent state to create a squeezed state, instead being pumped with a non-Gaussian state (e.g. depleted pump PDC~\cite{squeeze_pump_d}). This becomes a non-trivial interaction beyond the realm of Gaussian quantum mechanics, which can be approximately modelled with superpositions of Gaussian states. 

Another direction of post-Gaussian models involves a recent interest in hybrid models, involving combinations of discrete-variable (DV) systems (e.g. qubits) and continuous-variable (CV) systems (e.g. optical modes) in an effort to create a fault-tolerant quantum computer~\cite{qubit_beam_split}. This aims to exploit the existing toolbox of qubit gates and the long coherence times of optical modes for optimal performance. This comes at a cost as the presence of DV systems violates the Gaussianity of the system, allowing for quantum advantage but also complicating the modelling of such a system. 

Alternatively, a shifted perspective on hybrid models leads to qubit-controlled linear and quadratic displacements on CV systems. These non-Gaussian CV operations have been proposed for implementing universal hybrid quantum computing~\cite{trapped_ion_universal_QC}. Explorations of these techniques have suggested methods for qubit-controlled squeezing~\cite{trapped_ion_qubit_squeezing} and, more recently, have resulted in experimental preparation and manipulation of GKP states in trapped ion systems~\cite{matsos2024universalquantumgateset}.

In the realm of relativistic quantum information (RQI) a DV system as a control for linear and quadratic Gaussian unitaries on a CV system can be a useful tool for analogue gravity simulation~\cite{SineKG_LC_CIRCUIT,Bei_Lok_Charis_GR_Decoherence,Optomech_GR_decoherence}. Extending these controlled Gaussian unitaries beyond QO into QFT is mathematically well understood in the case of the linear displacements, especially the evaluation of operator expectation values from states generated by these controlled linear operators~\cite{Verdon-Akzam_QET,Funai_QET,Polo-Gomez_delta_qubit}. However, quadratic Gaussian unitaries controlled by qubits are less well understood, with most past works relating to few-mode systems and very specific types of quadratic couplings~\cite{two-mode_squeezed_superpositions}. Consequently, there is a gap in current theoretical models within the grasp of near-future experiments.

Formally, the textbook definition of a squeezed state tends to be written as $\ket{\phi_{\bm{K}}}=\exp[-\ii \hat{O}]\ket{0}$, where $\hat{O}=r \hat{a}^{\dagger}\hat{b}^{\dagger}+r^{*}\hat{a}\hat{b}$ (the notation describing quadratic Gaussian states is borrowed from~\cite{K_graphs_squeezed}). If $\hat{a}$ and $\hat{b}$ correspond to different modes (i.e. $[\hat{a},\hat{b}^{\dagger}]=0$) this is a two-mode squeezed state and we can use the Baker-Campbell-Hausdorff (BCH) method to find the Fock basis representation of the state~\cite{Gerry_knight}, including the corresponding global phase. These straightforward methods are used in~\cite{two-mode_squeezed_superpositions}, where the authors use the superposition of two-mode squeezed states to improve quantum sensing. In such a case, the phase of a squeezed state plays an important role and cannot be ignored. Our paper aims to generalise these results to arbitrary zero-mean, pure Gaussian states, i.e. where $\hat{O}$ is a general 2nd order polynomial in field operators, with no linear terms. In particular, the operator $\hat{O}$ includes terms of the form $\hat{a}^{\dagger}\hat{a}$, which appear in QO shearing operations but not squeezing. This is particularly necessary in QFT where the absence of $\hat{a}^{\dagger}\hat{a}$ terms often requires a spatially non-local operation, a significant problem when the rotating-wave approximation and single-mode approximation don't hold~\cite{EMM_causality_violations,Funai_rwa}.

The original problem that motivated this paper was in QFT, where the author attempted to modify the quantum energy teleportation (QET) protocol from a qubit-controlled linear field coupling ($\hat{\sigma}_{x}\otimes\hat{\phi}$)~\cite{Funai_QET} to a quadratic field coupling ($\hat{\sigma}_{x}\otimes\hat{\phi}^{2}$). The protocol was intentionally designed to permit non-perturbative modelling similar to Gaussian quantum mechanics. Unfortunately, this quadratic coupling falls beyond its simplifying reach, requiring tools capable of dealing with superpositions of Gaussian states and expectation values of such states. This paper presents a tool capable of resolving this issue, which has only recently been independently investigated~\cite{Hackl_superGaussian} for the case of quantum optics models.
% something that to our knowledge was an unasked and open problem.

The precise mathematical problem of superpositions of Gaussian states has 2 main forms, the resolution of either being sufficient to solve the physical problem stated above. Firstly, what is the overlap of two Gaussian states $\braket{\phi_{\bm{K}}|\phi_{\bm{K}'}}$ (including complex phase information)? Alternatively, given a Gaussian state in its unitary representation $\ket{\phi_{\bm{K}}}=\exp[-\ii \hat{O}]\ket{0}$ (quadratic unitary acting on the vacuum), what is its Fock basis representation (including global phase information). Both these problems are partially solved, although, neither of the solutions contains the necessary phase information to represent coherence in quantum superpositions.

In this paper we introduce a method of representing a general Gaussian state in a Fock basis representation whilst preserving the global phase information arising from its unitary representation. In section~\ref{sec_m_2} we overview relevant relativistic quantum information (RQI) and quantum optics (QO) definitions, given that this paper reports on a generally unexplored intersection of these two disciplines. In section~\ref{sec_m_3} we introduce quadratic Unruh-DeWitt detectors (UDW) and discuss the need for faithful Fock basis representations of Gaussian states. In section~\ref{sec_m_4} we present the main result of the paper, describing how to obtain the phase information that has been previously unknown when equating unitary and Fock representations of Gaussian states. In section~\ref{sec_m_5} we use the results presented in section~\ref{sec_m_4} to model a previously unsolved problem, namely how a detector's excitation probability is affected by an excited detector in its past (Fermi problem). This problem has been studied in the case of linear detectors, but never for quadratic detectors in a non-perturbative setting. Finally, in sections~\ref{sec_m_6} and~\ref{sec_m_7} we discuss the relevance and applicability of these results and present concluding remarks.

Note, in this manuscript we only consider zero-mean, pure Gaussian states. Generalisations to non-zero mean states can be made with minor modifications.
\section{Background}\label{sec_m_2}
\subsection{Interacting with a quantum field theory}
Unlike quantum optics models, quantum field theories are relativistically sensitive and therefore are incompatible with some of the simplifications implemented in quantum optical interactions, e.g. the single or few-mode approximation. One well-known issue in QFT involves projective measurements, and the divergences~\cite{Hu_2012} and causality violations it can introduce~\cite{sorkin1993,papageorgiou2023eliminating}.  As a result, interactions in quantum field theory are often mediated by a first quantised system called an Unruh-DeWitt (UDW) detector, usually a qubit that couples linearly to the field operator. 
\subsubsection*{Unruh-DeWitt detector}
The simplest UDW detector is a qubit that couples linearly to the field~\cite{UDW_1,UDW_2}, whose basic form is $\hat{\sigma}_{x} \hat{\phi}$. However, coupling to the field at a single point introduces divergences in quantum field theory, and consequently it is common to regularise the interaction by introducing a spatial smearing, i.e. giving the detector some effective non-zero size; and a switching function, which dictates when and how smoothly the detector is turned on and off. One particularly important difference between quantum field theory and quantum optics is that in QFT we do not implement any rotating wave approximation or Jaynes-Cummings type model, as excitation number preserving models violate causality and locality~\cite{EMM_causality_violations,Funai_rwa}. Generally, in QFTs, UDW detectors can be used to implement operations on the field and also perform weak measurements on the field. 

The simple UDW detector introduced above describes the ideal case of a qubit interacting with a scalar field, whereas the usual physical interaction between a first and second quantised system involves vector fields in the form of atomic light-matter interactions. Such an interaction can be derived from the $U(1)$ symmetry of an electron and under the dipole approximation can be written in the familiar form $\bm{r}\cdot\bm{E}$~\cite{Mayer_dipole}, where $\bm{r}$ is the position of the electron. Modelling this scenario requires more effort than the UDW model; although, when considering a situation with a single, dominant atomic transition, e.g. 1s$\rightarrow$2p in Hydrogen, the two models bear many similarities. In particular, the UDW model has been shown to be capable of capturing the physics of light-matter interactions whenever angular momentum is not exchanged~\cite{UDW_LM1,UDW_LM2}. The simple UDW model therefore eliminates the need to deal with vector fields or the additional microstructure of atomic detectors, whilst capturing the relevant physics.

Alternatively, motivated by Dirac fields and their U(1) symmetry, some studies have been conducted with quadratically coupled detectors~\cite{a_sachs_2UDW_1}, that is $\hat{\sigma}_{x} \hat{\phi}^{2}$. The primary motivation of this past study was to consider how detectors behave when coupled to a fermionic field and to study the differences between bosons and fermions. Recent papers have considered genuine quadratic coupling to bosonic fields~\cite{hummer_2UDW,a_sachs_2UDW_1}, especially as experimental proposals and experimental techniques begin to improve nonlinear light-matter coupling.

Mathematically, a quadratically coupled detector can come in several different forms. Sachs et al.~\cite{a_sachs_2UDW_1} considered a detector interaction of the form \mbox{$\int\dd\bm{x}\,F(\bm{x}):\hat{\phi}^2(\bm{x}):$}, where $F(\bm{x})$ is the spatial smearing of the detector and the colons $:\hat{\phi}^{2}(\bm{x}):$ corresponds to normal ordering. This is necessary to remove some divergences inherent in the interaction~\cite{hummer_2UDW}. More generally, we consider $\int\dd\bm{x}\dd\bm{y}\,F(\bm{x},\bm{y}):\phi(\bm{x})\phi(\bm{y}):$, which helps to remove additional divergences that appeared in past works. The normal ordering operation can be interpreted as a renormalization of the detector's Hamiltonian. Normal ordering is effective in removing divergences in 2nd order perturbations, whilst studies in higher order perturbation theory are lacking and it is not currently known if this suffices to regularise divergences in higher order perturbation theory.
\subsubsection*{$\delta$-coupling}
 When using UDW detectors in RQI it is usual to use perturbation theory, often 2nd order, and evaluate any relevant expression up to some order of the coupling constant. One of the few non-perturbative tools available in QFT is to consider an interaction where the switching function is a delta function, that is to say, the detector interacts very suddenly and very quickly for a brief instant of time. This technique has been used for past RQI protocols, including quantum energy teleportation~\cite{Hotta_2010}. It has also been studied as a possible alternative to BCH or Suzuki-Trotter methods of unitary evolution~\cite{Polo-Gomez_delta_qubit}.

Past works that have used this delta coupling have done so with the linear detector coupling, whereupon the post-interaction state consists of QFT coherent states entangled with the qubit detector. Namely,
\begin{align}
\hat{H}_{\textsc{i}}&=\lambda\delta(t)\hat{\sigma}_{x}(t)\int\dd^{3}\bm{x}\,F(\bm{x})\hat{\phi}(t,\bm{x}),\\
\hat{U}\ket{0}&(a\ket{\px}+b\ket{\mx})=a\ket{\bm{\alpha}}\ket{\px}+b\ket{-\bm{\alpha}}\ket{\mx},
\end{align}
where $\ket{0}$ is the vacuum state of the QFT and $\hat{U}$ is the unitary generated by $\hat{H}_{\textsc{i}}$.
In quantum optical terms this can be viewed as a linear coherent displacement of the field with a qubit control. What makes this technique useful and solvable are the well-known relations describing the union of concatenated displacement operators as well as the description of a coherent state in the Fock number basis. In QFT, these displacement operator relations are:
\begin{align}
\hat{D}(\bm{\alpha})&=\exp\left[\int\dd^{3}\bm{k}\left(\alpha_{\bm{k}}^{\vphantom{*}}\hat{a}_{\bm{k}}^{\dagger}-\alpha_{\bm{k}}^{*}\hat{a}_{\bm{k}}^{\vphantom{\dagger}}\right)\right],\\
\hat{D}(\bm{\alpha})\hat{D}(\bm{\beta})&=\exp\left[\frac{1}{2}\int\dd^{3}\bm{k}\left(\alpha_{\bm{k}}^{\vphantom{*}}\beta_{\bm{k}}^{*}-\alpha_{\bm{k}}^{*}\beta_{\bm{k}}^{\vphantom{*}}\right)\right]\nonumber\\
&\qquad\times\hat{D}(\bm{\alpha}+\bm{\beta}),\label{eq4}\\
\hat{D}(\bm{\alpha})\ket{0}&=e^{-\frac{1}{2}\|\bm{\alpha}\|}\exp\left[\int\dd^{3}\bm{k}\,\alpha_{\bm{k}}^{\vphantom{*}}\hat{a}_{\bm{k}}^{\dagger}\right]\ket{0},\label{eq5}\\
\|\bm{\alpha}\|&=\int\dd^{3}\bm{k}\,\abs{\alpha^{\vphantom{*}}_{\bm{k}}}^{2}.
\end{align}
The equivalent set of relations for quadratic displacements will be needed to model quadratic detectors with $\delta$-couplings.
\subsection{Quadratically generated unitaries in QFT}
In the case of quadratic, $\delta$-coupled detectors, the initial calculations are very similar to those of the linear case. The post-interaction state can be described as a Gaussian unitary controlled by a qubit. Unfortunately, quadratic Gaussian unitaries do not have simple fusion rules, such as those of linear displacement operators \eqref{eq4}. 
\subsubsection*{Nullifier equation}
Quadratic Gaussian unitaries present mathematical issues when using the BCH formula to derive fusion rules analogous to \eqref{eq4}. Instead, it is more practical to describe a zero-mean Gaussian state as an element of the kernel of some linear field operator~\cite{K_graphs_squeezed,gauss_k_graph}. Consider an invertible matrix $\bm{G}_{0}$, a quadratic Gaussian $\hat{S}$ and the vector of annihilation operators $\hat{\bm{a}}$:
\begin{align}
\hat{S}\left(\bm{G}_{0}\hat{\bm{a}}\right)\ket{0}&=\bm{0},\\
\hat{S}\left(\bm{G}_{0}\hat{\bm{a}}\right)\hat{S}^{\dagger}\hat{S}\ket{0}&=\bm{0},\\
\left(\bm{G}_{1}\hat{\bm{a}}+\bm{G}_{2}\hat{\bm{a}}^{\dagger}\right)\hat{S}\ket{0}&=\bm{0},\\
\hat{\bm{L}}\hat{S}\ket{0}&=\bm{0},
\end{align}
$\bm{G}_{1}$ and $\bm{G}_{2}$ are matrices arising from the conjugation properties of $\hat{S}$. The operator $\hat{\bm{L}}=\left(\bm{G}_{1}\hat{\bm{a}}+\bm{G}_{2}\hat{\bm{a}}^{\dagger}\right)$ is often referred to as the nullifier~\cite{K_graphs_squeezed} (i.e. maps the Gaussian state to zero). The nullifier equation is described in appendix~\ref{app_sec_a} in more detail. Furthermore, if $\hat{S}$ consists of a product of quadratic Gaussian unitaries, the operator $\hat{\bm{L}}$ can still be evaluated using the above conjugation approach with relative ease. 

The set of all nullifiers $\{\hat{\bm{L}}\}$ can be spanned by the following basis representation:
\begin{align}
\left(\hat{\bm{a}}-\bm{\Delta k}\bm{K}\hat{\bm{a}}^{\dagger}\right)_{\bm{k}} \hat{S}\ket{0}&=0,\label{eq13}
\end{align}
where $\bm{K}$ is a symmetric matrix. This equation is called the nullifier equation and is used to determine the appropriate $\bm{K}$ matrix for nullifying $\hat{S}\ket{0}$~\cite{K_graphs_squeezed}. This is from where we derive our notation: $\ket{\phi_{\bm{K}}}=\hat{S}\ket{0}$ is the quadratic Gaussian satisfying \eqref{eq13} above.  The matrix $\bm{K}$ is determined by $\hat{S}$ up to a phase, i.e. $\bm{K}[\hat{S}]=\bm{K}[e^{\ii\phi}\hat{S}]$.  Note that $\bm{\Delta k}$ is the momentum space Riemann sum measure, which is present since we are considering quantum fields, where the canonical commutation relations are Dirac delta normalised (c.f. quantum optics and Kronecker delta normalised models; see appendix~\ref{sec_a_b} for details).

The primary advantage of the nullifier equation is that it allows one to easily represent a Gaussian state in a Fock basis. By considering the state's series expansion in the Fock basis, we can solve the nullifier equation and find that
\begin{align}
\hat{S}\ket{0}&=\mathcal{N} e^{\bm{\Delta k}^{2}\frac{\hat{\bm{a}}^{\textsc{h}}\bm{K}\hat{\bm{a}}^{\dagger}}{2}}\ket{0},\label{eqi8}
\end{align}
where $\mathcal{N}\in\mathbb{C}$ is a normalisation factor, $\hat{\bm{a}}^{\dagger}$ is the column vector consisting of creation operators and $\hat{\bm{a}}^{\textsc{h}}$ is the row vector consisting of creation operators, i.e. daggered ladder operators and transposed column vector. The Fock state represented above uniquely solves the nullifier equation, up to the normalisation factor $\mathcal{N}$, whose phase cannot be determined from this approach. The following sections of this paper derive a procedure that can be used to evaluate the phase corresponding to the exact $\hat{S}$ operator.

For those familiar with Bogoliubov transformations, the equation above should inspire some familiarity as the operator acting on the Gaussian state \eqref{eq13} is an annihilation operator that has undergone a Bogoliubov transformation~\cite{Bogoljubov1958,RevModPhys.77.513}. From this interpretation, it is common to refer to the state in \eqref{eqi8} as the squeezed vacuum, that is to say, a state that can be defined as the vacuum of a set of annihilation operators that have undergone a Bogoliubov transformation. Understanding how to properly manipulate these Gaussian states is one of the tools that will arise as we consider, for example, superpositions of space-time metrics and the different vacuums that they induce.  

In this manuscript, we will focus our attention on zero-mean Gaussian states only. Gaussian states with non-zero mean can be represented as a linear displacement on a zero-mean state and therefore can be easily dealt with using the results presented here.
\section{Quadratic detectors in cavity fields}\label{sec_m_3}
In a cavity a detector’s mathematical form is very similar to that of a detector outside of a cavity, with the added caveat that the detector must be small with respect to the cavity~\cite{Funai_thesis}. This may seem like an obvious requirement, but it is important to remember that we are considering a cavity that is sensitive to relativistic dynamics, and of course, the detector is considered to be a first quantized system and capable of instant communication within itself. Detectors are often modelled as atom-sized, and this merely puts restrictions on our cavity to be significantly larger than the diameter of an atom, a thoroughly reasonable and generally respected condition. Main differences associated with a detector in a cavity tend to be mathematical and, in particular, there is a simplifying difference due to the introduction of an IR cut-off, which has the effect of discretising the momentum spectrum of the mode decomposition (see Appendix~\ref{sec_a_b} for details). In this manuscript we only consider systems of 3 spatial dimensions.
\subsubsection*{General quadratic interactions}
The most general quadratic detector interacting with the field is given by the following Hamiltonian 
\begin{align}
\hat{H}(t)&=\lambda\chi(t)\hat{\sigma}_{x}(t)\int\dd \bm{x}\dd\bm{y}\,\left[F(\bm{x},\bm{y}) :\hat{\phi}(\bm{x})\hat{\phi}(\bm{y}):\right.\nonumber\\
\,\,&+
G(\bm{x},\bm{y}) \left(:\hat{\pi}(\bm{x})\hat{\phi}(\bm{y})+\hat{\phi}(\bm{x})\hat{\pi}(\bm{y}):\right)\nonumber\\
\,\,&+\left.P(\bm{x},\bm{y}) :\hat{\pi}(\bm{x})\hat{\pi}(\bm{y}):
\vphantom{F(\bm{x},\bm{y}) :\hat{\phi}(\bm{x})\hat{\phi}(\bm{y}):}\right],\label{eqb9}
\end{align} 
where the colons $:\hat{O}:$ indicate normal ordering, which we introduce to avoid some divergences associated with this interaction~\cite{a_sachs_2UDW_1,hummer_2UDW}. Since we intend to employ $\chi(t)=\delta(t)$ switching functions, the interaction Hamiltonian above has the field operators in the Schr\"{o}dinger picture, which helps simplify the description and calculations of future protocols. The interaction above is the most general quadratic detector interaction that is local, i.e. a detector that is sensitive only on the support of $F,G$ and $P$. To simplify future calculations, we rewrite \eqref{eqb9} as
\begin{align}
\hat{H}(t)&=\lambda\delta(t)\hat{\sigma}_{x}(t)\otimes\hat{h},
\end{align}
where $\hat{h}$ is a generator of quadratic displacements in the quantum field and we have also stipulated that the switching function is a Dirac $\delta$-function. The quadratic generator can also be written in a ladder operator basis as
\begin{align}
\hat{h}&=\bm{\Delta k}^{2}\sum_{\bm{kk'}}A_{\bm{kk'}}\hat{a}_{\bm{k}}^{\vphantom{\dagger}}\hat{a}_{\bm{k}'}^{\vphantom{\dagger}}
+2B_{\bm{kk'}}\hat{a}_{\bm{k}}^{\dagger}\hat{a}_{\bm{k}'}^{\vphantom{\dagger}}
+\conjnf{A}_{\bm{kk'}}\hat{a}_{\bm{k}}^{\dagger}\hat{a}_{\bm{k}'}^{\dagger},\label{eq17}
\end{align}
where the matrices $\bm{A}$ and $\bm{B}$ are given by
\begin{align}
A_{\bm{kk'}}&=\mathcal{F}_{\bm{kk'}}+\ii\mathcal{G}_{\bm{kk'}}+\ii\mathcal{G}_{\bm{k'k}}-\mathcal{P}_{\bm{kk'}},\label{eq150}\\
B_{\bm{kk'}}&=\mathcal{F}_{\bm{kk'}}+\ii\mathcal{G}_{\bm{kk'}}-\ii\mathcal{G}_{\bm{k'k}}+\mathcal{P}_{\bm{kk'}},\label{eq151}
\end{align}
with
\begin{align}
\mathcal{F}_{\bm{kk'}}&=\frac{4}{\pi^{3}\sqrt{\omega\omega'}}\int\dd\bm{x}\dd\bm{y}\,\bm{F}(\bm{x},\bm{y})\prod_{i=1}^{3}\sin(k_{i}x^{i})\sin(k_{i}'y^{i}),\label{eq_z_15}\\
\mathcal{G}_{\bm{kk'}}&=-\frac{4}{\pi^{3}}\sqrt{\frac{\omega'}{\omega}}\int\dd\bm{x}\dd\bm{y}\,\bm{G}(\bm{x},\bm{y})\prod_{i=1}^{3}\sin(k_{i}x^{i})\sin(k_{i}'y^{i}),\\
\mathcal{P}_{\bm{kk'}}&=\frac{4\sqrt{\omega\omega'}}{\pi^{3}}\int\dd\bm{x}\dd\bm{y}\,\bm{P}(\bm{x},\bm{y})\prod_{i=1}^{3}\sin(k_{i}x^{i})\sin(k_{i}'y^{i}),\label{eq_z_17}
\end{align}
 where the spatial integral's domain is the interior of the cavity. Note, in this manuscript we consider massless fields only, i.e. $\omega=\abs{\bm{k}}$.
 
What we have in \eqref{eq17} is a general quadratic Hamiltonian for a quantum field, where the matrices $\bm{A}$ and $\bm{B}$ are symmetric and hermitian respectively, and also restricted to ensure they describe a spatially local operation. Note that we are considering rectangular cavities, hence, $k_{i}=n_{i}\pi/L_{i}$ is the wavenumber of the mode decomposition, where $n_{i}\in\mathbb{Z}$ and $L_{i}$ is the length of the $i$th dimension of the cavity. This defines a countably infinite set, however, a UV cut-off can be introduced if finite matrices are required.

The quadratic Hamiltonian in \eqref{eq17} is the general form that we consider for generating zero-mean Gaussian states. Importantly, this expression contains excitation-preserving quadratic terms (i.e. $\hat{a}^{\dagger}\hat{a}$), even when written in a Jordan-normal-form mode basis. Consequently, we cannot use the BCH method to find a closed-form expression of the global phase term in the Fock representation. In the sections below, we evaluate the nullifier equation corresponding to this quadratic Hamiltonian and then determine the phase of $\mathcal{N}$ necessary to equate the Fock representation of a Gaussian state with its unitary representation. 
\subsubsection*{Constructing a nullifier equation}
To proceed, we need to find the Fock representation of the Gaussian state
\begin{align}
\hat{S}(\lambda)\ket{0}&=e^{-\ii\lambda\hat{h}}\ket{0},\label{eq23}
\end{align}
whose first step requires determining its corresponding nullifier equation. To determine the exact form of this equation, we exploit the conjugation relations of quadratic Gaussian unitaries. That is: (see details in appendix~\ref{app_sec_a}) 
\begin{align}
\hat{S}(-\lambda)\left(\bm{G}_{1}\hat{\bm{a}}+\bm{G}_{2}\hat{\bm{a}}^{\dagger}\right)&\hat{S}(\lambda)=\begin{pmatrix}
\bm{G}_{1} & \bm{G}_{2}
\end{pmatrix}\nonumber\\
\times&\text{exp}\left[2\ii\lambda\bm{\Delta k}
\begin{pmatrix}
-\bm{B} & -\conjnf{\bm{A}} \\
\bm{A} & \conjnf{\bm{B}}
\end{pmatrix}\right]
\begin{pmatrix}
\hat{\bm{a}}\\
\hat{\bm{a}}^{\dagger}
\end{pmatrix},\label{eq24}
\end{align}
where $\bm{G}_{i}$ are arbitrary square matrices. Also note, $\hat{S}(-\lambda)=\hat{S}^{\dagger}(\lambda)$. From the equation above, we can see that a careful choice of matrices $\bm{G}_{i}$ will result in the right-hand side consisting exclusively of annihilation operators. Therefore, by choosing
\begin{align}
\begin{pmatrix}
\bm{G}_{1} & \bm{G}_{2}
\end{pmatrix}&=
\begin{pmatrix}
\mathbb{I} & \bm{0}
\end{pmatrix}
\text{exp}\left[-2\ii\lambda\bm{\Delta k}
\begin{pmatrix}
-\bm{B} & -\conjnf{\bm{A}} \\
\bm{A} & \conjnf{\bm{B}}
\end{pmatrix}\right],\label{eqb19}
\end{align}
then combining \eqref{eq23} and \eqref{eq24} we obtain a nullifier equation:
\begin{align}
\left(\bm{G}_{1}\hat{\bm{a}}+\bm{G}_{2}\hat{\bm{a}}^{\dagger}\right)\hat{S}(\lambda)\ket{0}&=0.
\end{align}
Comparing this equation with \eqref{eq13} we can see that
\begin{align}
\bm{\Delta k}\bm{K}&=-\bm{G}_{1}^{-1}\bm{G}_{2}^{\vphantom{-1}}.
\end{align}
We therefore have the nullifier equation corresponding to the Gaussian state generated by a general local quadratic interaction with $\delta(t)$-switching.

\begin{widetext}

\section{Evaluating the phase of the Fock representation}\label{sec_m_4}
As mentioned above, the objective of this paper is to evaluate the phase between the unitary representation of a zero-mean Gaussian state and the Fock representation of said state. It is sufficient to compare the coefficients of the zero Fock vector $\ket{0}$ of the respective representations, which boils down to the vacuum expectation value of the quadratic Gaussian unitary. The equivalence of the determination of the phase and the evaluation of the vacuum expectation value of the quadratic Gaussian unitary suggests an approach to solving this problem.

 The vacuum expectation value of the quadratic Gaussian unitary is not known and can only be evaluated perturbatively. Our idea is to find a non-trivial expression that reveals the phase of the Fock representation while not involving the vacuum expectation value of the quadratic Gaussian unitary. To that end, consider the Gaussian state \eqref{eq23} parameterized by the interaction strength $\lambda$. This state has two representations, c.f.~\eqref{eqi8}:
\begin{align}
\hat{S}(\lambda)\ket{0}&=e^{-\ii\lambda \hat{h}}\ket{0}=e^{D}e^{\bm{\Delta k}^{2}\frac{\hat{\bm{a}}^{\textsc{h}}\bm{K}\hat{\bm{a}}^{\dagger}}{2}}\ket{0},
\end{align}
where $D\in\mathbb{C}$, $\hat{\bm{a}}^{\dagger}$ is the column vector consisting of creation operators and $\hat{\bm{a}}^{\textsc{h}}$ is the row vector consisting of creation operators, i.e. daggered ladder operators and transposed column vector. The real part of $D$ is known since the state is normalised and $\mathcal{N}=e^{D}$, c.f. \eqref{eqi8}. In order to find the imaginary part of $D$, i.e. the phase of the Fock representation, we take a derivative with respect to $\lambda$ and then take the inner product with the same Gaussian state:
\begin{align}
\bra{0}\hat{S}^{\dagger}(\lambda)\partial_{\lambda}\hat{S}(\lambda)\ket{0}&=
\bra{0}e^{\ii\lambda \hat{h}}\left(-\ii\hat{h}\right)e^{-\ii\lambda\hat{h}}\ket{0},\\
&=\bra{0}\left(-\ii\hat{h}\right)\ket{0}=0,\label{eq31}
\end{align}
where we used the fact that $\hat{h}$ commutes with the unitary it generates. Note that \eqref{eq31} evaluates to zero since $\hat{h}$ is normal-ordered. This is unsurprising given that this Gaussian state is a unit-length vector in a Hilbert space. In the Fock representation:
%\begingroup
%\allowdisplaybreaks[0]
\begin{align}
\bra{0}\hat{S}^{\dagger}(\lambda)\partial_{\lambda}\hat{S}(\lambda)\ket{0}=
\bra{0}e^{\bm{\Delta k}^{2}\frac{\hat{\bm{a}}^{\textsc{t}}\conjnf{\bm{K}}\hat{\bm{a}}}{2}}e^{\conjnf{D}}\left(D'+\bm{\Delta k}^{2}\frac{\hat{\bm{a}}^{\textsc{h}}\bm{K}'\hat{\bm{a}}^{\dagger}}{2}\right)
e^{D}e^{\bm{\Delta k}^{2}\frac{\hat{\bm{a}}^{\textsc{h}}\bm{K}\hat{\bm{a}}^{\dagger}}{2}}\ket{0}=0,
\end{align}
%\endgroup
where the prime indicates a derivative in $\lambda$. Unlike the unitary representation, this expression is more complicated and requires some work. Since the exact form of $\bm{K}$ is known, we can evaluate its derivative and solve for $D'$. We can now see the advantage of this approach, i.e. we have obtained an expression for $D$ without the need to evaluate the vacuum expectation value of a quadratic Gaussian unitary.

Unfortunately, the expression above is anti-normal ordered, i.e. there are annihilation operators on the left and creation operators on the right. Since we can't easily commute these operators, we instead introduce a resolution of the identity in terms of coherent states:
\begin{align}
0&=\bra{0}e^{D+\conjnf{D}} e^{\bm{\Delta k}^{2}\frac{\hat{\bm{a}}^{\textsc{t}}\conjnf{\bm{K}}\hat{\bm{a}}}{2}}\,
\mathbb{I}\left(D'+\bm{\Delta k}^{2}\frac{\hat{\bm{a}}^{\textsc{h}}\bm{K}'\hat{\bm{a}}^{\dagger}}{2}\right)
e^{\bm{\Delta k}^{2}\frac{\hat{\bm{a}}^{\textsc{h}}\bm{K}\hat{\bm{a}}^{\dagger}}{2}}\ket{0},\\
&=\bra{0}e^{D+\conjnf{D}} e^{\bm{\Delta k}^{2}\frac{\hat{\bm{a}}^{\textsc{t}}\conjnf{\bm{K}}\hat{\bm{a}}}{2}}
\prod_{i}\frac{\bm{\Delta k}}{\pi}\int\dd^{2}\alpha_{i}\,\ketbra{\bm{\alpha}}{\bm{\alpha}}
\left(D'+\bm{\Delta k}^{2}\frac{\hat{\bm{a}}^{\textsc{h}}\bm{K}'\hat{\bm{a}}^{\dagger}}{2}\right)
e^{\bm{\Delta k}^{2}\frac{\hat{\bm{a}}^{\textsc{h}}\bm{K}\hat{\bm{a}}^{\dagger}}{2}}\ket{0},\label{eqt29}\\
&=\prod_{i}\left(\frac{\bm{\Delta k}}{\pi}\right) e^{D+\conjnf{D}}\int\dd^{2}\bm{\alpha}\,\left(D'+\bm{\Delta k}^{2}\frac{\bm{\alpha}^{\dagger}\bm{K}'\conjnf{\bm{\alpha}}}{2}\right)\text{exp}\left[\bm{\Delta k}^{2}
\frac{\bm{\alpha}^{\textsc{t}}\conjnf{\bm{K}}\bm{\alpha}+
\bm{\alpha}^{\dagger}\bm{K}\conjnf{\bm{\alpha}}}{2}-\bm{\Delta k}\|\bm{\alpha}\|^{2}
\right],
\end{align}
where $\ket{\bm{\alpha}}$ is a QFT coherent state \eqref{eq5} and
\begin{align}
\|\bm{\alpha}\|^{2}=\sum_{i=1}^{n}\abs{\alpha_{i}}^{2}.
\end{align}
By evaluating the action of the ladder operators on coherent states, we now have an expression that is free of operators and states. To evaluate this integral, define
\begin{align}
\bm{\aleph}&=\begin{pmatrix}
2\bm{\Delta k}-\bm{\Delta k}^{2}\left(\bm{K}+\conjnf{\bm{K}}\right) & \ii\bm{\Delta k}^{2}(\bm{K}-\conjnf{\bm{K}}) \\
\ii\bm{\Delta k}^{2}(\bm{K}-\conjnf{\bm{K}}) & 2\bm{\Delta k}+\bm{\Delta k}^{2}\left(\bm{K}+\conjnf{\bm{K}}\right)
\end{pmatrix},\\
\bm{\aleph}^{-1}&=\begin{pmatrix}
\bm{\Gamma} & \bm{\Sigma}\\
\bm{\Xi} & \bm{\Psi}
\end{pmatrix},
\end{align}
then the Gaussian integral takes the form
\begin{align}
0&=\prod_{i}\left(\frac{\bm{\Delta k}}{\bm{\pi}}\right)e^{D+\conjnf{D}}
\int\dd\bm{x}\dd \bm{y}\,\text{exp}\left[
-\frac{1}{2}
\begin{pmatrix}
\bm{x}^{\textsc{t}} & \bm{y}^{\textsc{t}}
\end{pmatrix}
\bm{\aleph}
\begin{pmatrix}
\bm{x}\\
\bm{y}
\end{pmatrix}
\right]
\left(
\frac{D'}{\bm{\Delta k}^{2}}+\bm{x}^{\textsc{t}}\frac{\bm{K}'}{2}\bm{x}
-\ii\bm{x}^{\textsc{t}}\frac{\bm{K}'}{2}\bm{y}
-\ii\bm{y}^{\textsc{t}}\frac{\bm{K}'}{2}\bm{x}
-\bm{y}^{\textsc{t}}\frac{\bm{K}'}{2}\bm{y}
\right),
\end{align}
where $\bm{\alpha}=\bm{x}+\ii\bm{y}$. This Gaussian integral can be evaluated and simplified into
\begin{align}
0&=D'+\frac{\bm{\Delta k}^{2}}{2}\Tr\left(\left[\bm{\Gamma}-\ii\bm{\Xi}-\ii\bm{\Sigma}-\bm{\Psi}\right]\bm{K}'\right),\label{eq40}
\end{align}
where $\prod_{i}\frac{\bm{\Delta k}}{\bm{\pi}}e^{D+\conjnf{D}}$ can be cancelled out as the LHS equals zero.
 
In \eqref{eq40} we now have an expression for the phase of the Fock representation of a Gaussian state. The central idea behind this technique was to introduce a derivative in $\lambda$ that breaks the symmetry between the state (ket) and its dual (bra). This can be extended to an arbitrary number of sequential quadratic displacements, although the complexity of the problem is soon unwieldy. Note that this expression does not require the use of the BCH formula or any such concatenation of commutation relations. That being said, this equation is not always easy to solve. 
\subsection{$\phi^{2}$ example}\label{sec_phi2}
If we consider the simple case of a $\phi^{2}$ interaction (i.e. $G=P=0$ in \eqref{eqb9}), we can simplify the nullifier equation and evaluate the $\bm{K}$ matrix exactly. From \eqref{eqb19} we can evaluate $\bm{K}$:
\begin{align}
\begin{pmatrix}
\bm{G}_{1} & \bm{G}_{2}
\end{pmatrix}
&=
\begin{pmatrix}
\mathbb{I} & \bm{0}
\end{pmatrix}\text{exp}\left[-2\ii\lambda\bm{\Delta k}
\begin{pmatrix}
-\bm{\mathcal{F}} & -\bm{\mathcal{F}}\\
\bm{\mathcal{F}} & \bm{\mathcal{F}}
\end{pmatrix}
\right],\\
&=\begin{pmatrix}
\mathbb{I}+2\ii\lambda\bm{\Delta k} \bm{\mathcal{F}} & 2\ii\lambda\bm{\Delta k}\bm{\mathcal{F}}
\end{pmatrix}.
\end{align}
Therefore
\begin{align}
\bm{\Delta k}\bm{K}&=-\bm{G}_{1}^{-1}\bm{G}_{2}^{\vphantom{-1}},\\
&=-\left(\mathbb{I}+2\ii\lambda\bm{\Delta k}\bm{\mathcal{F}}\right)^{-1}2\ii\lambda\bm{\Delta k}\bm{\mathcal{F}},\\
\bm{K}&=-\left(\mathbb{I}+2\ii\lambda\bm{\Delta k}\bm{\mathcal{F}}\right)^{-1}2\ii\lambda\bm{\mathcal{F}}.
\end{align}
Continuing to evaluate all the relevant matrices:
\begin{align}
\bm{\aleph}&=\bm{\Delta k}\begin{pmatrix}
4\mathbb{I}-2\left(\mathbb{I}+4\lambda^{2}\bm{\Delta k}^{2}\bm{\mathcal{F}}^{2}\right)^{-1} & 4\lambda\bm{\Delta k}\bm{\mathcal{F}}\left(\mathbb{I}+4\lambda^{2}\bm{\Delta k}^{2}\bm{\mathcal{F}}^{2}\right)^{-1} \\
4\lambda\bm{\Delta k}\bm{\mathcal{F}} \left(\mathbb{I}+4\lambda^{2}\bm{\Delta k}^{2}\bm{\mathcal{F}}^{2}\right)^{-1} & 2\left(\mathbb{I}+4\lambda^{2}\bm{\Delta k}^{2}\bm{\mathcal{F}}^{2}\right)^{-1}
\end{pmatrix}.
\end{align}
Fortunately, we can also determine the inverse of this matrix too:
\begin{align}
\bm{\aleph}^{-1}&=\frac{1}{\bm{\Delta k}}
\begin{pmatrix}
\frac{\mathbb{I}}{2} & -\lambda\bm{\Delta k}\bm{\mathcal{F}}\\
-\lambda\bm{\Delta k}\bm{\mathcal{F}} & \frac{\mathbb{I}}{2}+4\lambda^{2}\bm{\Delta k}^{2}\bm{\mathcal{F}}^{2}
\end{pmatrix}.
\end{align}
Combining these results:
\begin{align}
D'&=-\frac{\bm{\Delta k}}{2}\Tr\left(\left[\frac{\mathbb{I}}{2}+\ii\lambda\bm{\Delta k}\bm{\mathcal{F}}+\ii\lambda\bm{\Delta k}\bm{\mathcal{F}}-\frac{\mathbb{I}}{2}-4\lambda^{2}\bm{\Delta k}^{2}\bm{\mathcal{F}}^{2}\right]\bm{K}'\right),\\
&=-\frac{\bm{\Delta k}}{2}\Tr\left(\left[2\ii\lambda\bm{\Delta k}\bm{\mathcal{F}}-4\lambda^{2}\bm{\Delta k}^{2}\bm{\mathcal{F}}^{2}\right]\bm{K}'\right),
\end{align}
and finally
\begin{align}
D'&=-\frac{\bm{\Delta k}}{2}\Tr\left(2\ii\lambda\bm{\Delta k}\bm{\mathcal{F}}\left[\mathbb{I}+2\ii\lambda\bm{\Delta k}\bm{\mathcal{F}}\right]\bm{K}'\right),\\
&=-\frac{\bm{\Delta k}}{2}\Tr\left(2\ii\lambda\bm{\Delta k}\bm{\mathcal{F}}(-1)2\ii\bm{\mathcal{F}}(\mathbb{I}+2\ii\lambda\bm{\Delta k}\bm{\mathcal{F}})^{-1}\right),\\
&=-\Tr\left(\frac{2\lambda\bm{\Delta k}^{2}\bm{\mathcal{F}}^{2}}{\mathbb{I}+2\ii\lambda\bm{\Delta k}\bm{\mathcal{F}}}\right),
\end{align}
where we can use the fraction notation as the numerator and denominator commute. This matrix equation can be integrated:
\begin{align}
D&=\ii\lambda\Tr(\bm{\Delta k}\bm{\mathcal{F}})-\frac{1}{2}\Tr(\ln(\mathbb{I}+2\ii\lambda\bm{\Delta k}\bm{\mathcal{F}})).\label{eq52}
\end{align}
Taking the trace of $\bm{\mathcal{F}}$ is a relatively simply calculation, although evaluating $\ln(\mathbb{I}+2\ii\lambda\bm{\mathcal{F}})$ requires knowledge of the specific eigenvalues of $\bm{\mathcal{F}}$.

Recall that
\begin{align}
\mathcal{F}_{\bm{kk'}}&=\frac{4}{\pi^{3}\sqrt{\omega\omega'}}\int\dd\bm{x}\dd\bm{y}\,\bm{F}(\bm{x},\bm{y})\prod_{i=1}^{3}\sin(k_{i}x^{i})\sin(k_{i}'y^{i}).
\end{align}
Finding the eigenvalues of $\bm{\mathcal{F}}$ is non-trivial due to the $1/\sqrt{\omega\omega'}$ prefactors. Mathematically, the problem of phase determination has shifted from evaluations of concatenated commutation relations and Lie algebra manipulation to evaluating transcendental functions of matrices. 
The simplest case for evaluation occurs when the smearing function is separable, i.e. $F(\bm{x},\bm{y})=f(\bm{x})f(\bm{y})$. In this case, the $\bm{\mathcal{F}}$ matrix becomes rank one. Consequently, when evaluating the logarithm of this matrix in \eqref{eq52}, we only need to consider a single eigenvalue and thus the expression can be evaluated exactly. Obviously, the higher the rank of the $\bm{\mathcal{F}}$ matrix, the greater the difficulty of this calculation.  

From the general form of the phase in~\eqref{eq52} we can now take the formal mathematical limit of $L_{i}\rightarrow\infty$ to generalise the result to non-cavity field situations. Physically, this limit should have virtually no effect provided the field perturbations we are modelling are far from the walls, i.e. the support of $F(\bm{x},\bm{y})$ is near $(L_{1}/2,L_{2}/2,L_{3}/2)$. In such a case, the presence of cavity walls should induce small, insignificant effects that vanish in the $L_{i}\rightarrow\infty$ limit~\cite{Funai_thesis}.
\section{Fermi problem with quadratic detectors}\label{sec_m_5}
Once we have this tool at our disposal, we can think of problems where this technique is necessary. The simplest of these problems is the so-called Fermi problem~\cite{Fermi_RWA}, which is also sometimes used as a tool for evaluating communication in quantum field theory~\cite{EMM_causality_violations}. The problem involves two detectors, one of which is situated in the past of the other according to some fixed observer. The first detector, Alice, will try to communicate a classical bit where `0' corresponds to not interacting with the field ($\lambda_{\textsc{a}}=0$) and `1' corresponds to interacting with the field ($\lambda_{\textsc{a}}\neq 0$). The second detector, Bob, will then interact with the field and try to learn what Alice's bit was. Naturally, if Alice and Bob are space-like separated, causality dictates that Bob is unable to evaluate Alice's bit. We consider this problem under the new conditions that Alice and Bob interact with the field using a quadratic detector and both using a Dirac delta switching function. This problem contains several of the mathematical techniques that we anticipate will be necessary in future calculations, and is also a problem that cannot be solved if one does not know the exact value of the phase $D$. 

Mathematically, the Fermi problem has an interaction Hamiltonian of the form
\begin{align}
\hat{H}_{\textsc{i}}&=\lambda_{\textsc{a}}\delta(t)\hat{\sigma}_{x,\textsc{a}}\int\dd\bm{x}\dd\bm{y}\,F_{\textsc{a}}(\bm{x},\bm{y})
:\hat{\phi}(\bm{x})\hat{\phi}(\bm{y}):
+\lambda_{\textsc{b}}\delta(t-T)\hat{\sigma}_{x,\textsc{b}}\int\dd\bm{x}\dd\bm{y}\,F_{\textsc{b}}(\bm{x},\bm{y})
:\hat{\phi}(\bm{x})\hat{\phi}(\bm{y}):.
\end{align}
The non-interacting portion of the Hamiltonian is the usual free-field and qubit evolution Hamiltonian. Note, the $\delta(t)$ coupling means we don't need to know the qubit energy gaps, only the qubits' state at the time of interaction. The field is initially taken to be in the vacuum state, whilst Alice and Bob's qubits are represented by:
\begin{align}
\ket{\psi_{\textsc{a}}(0^{-})}&=a_{1}\ket{\px}+a_{2}\ket{\mx},\label{eq_ab_54}\\
\ket{\psi_{\textsc{b}}(T^{-})}&=b_{1}\ket{\px}+b_{2}\ket{\mx},\label{eq_ab_55}
\end{align}
where $\ket{\pm}$ is the Pauli x-basis and the states are defined at times $t=0^{-},T^{-}$ for mathematical convenience.

The $\delta$-coupling means we can decompose the dynamics into 3 separate time intervals: Alice's interaction, free field evolution, and Bob's interaction. The system's state slightly before and after Alice's and Bob's interactions can be written as:
\begin{align}
\ket{\Psi(0^{-})}&=\ket{\psi_{\textsc{b}}(0)}\otimes\left(a_{1}\ket{\px}_{\textsc{a}}\ket{0}+a_{2}\ket{\mx}_{\textsc{a}}\ket{0}\right),\\
\ket{\Psi(0^{+})}&=\ket{\psi_{\textsc{b}}(0)}\otimes\left(a_{1}\ket{\px}_{\textsc{a}}\hat{S}_{\textsc{a}}(\lambda^{\textsc{a}})\ket{0} +a_{2}\ket{\mx}_{\textsc{a}}\hat{S}_{\textsc{a}}(-\lambda^{\textsc{a}})\ket{0}\right),\\
\ket{\Psi(T^{-})}&=\left(b_{1}\ket{\px}_{\textsc{b}}+b_{2}\ket{\mx}_{\textsc{b}}\right)\otimes\left(a_{1}\ket{\px;T}_{\textsc{a}}\hat{S}_{\textsc{a}}(\lambda^{\textsc{a}};T)\ket{0}+a_{2}\ket{\mx;T}_{\textsc{a}}\hat{S}_{\textsc{a}}(-\lambda^{\textsc{a}};T)\ket{0}\right),\\
\begin{split}
\ket{\Psi(T^{+})}&=b_{1}\ket{\px}_{\textsc{b}}\left(a_{1}\ket{\px;T}_{\textsc{a}}\hat{S}_{\textsc{b}}(\lambda^{\textsc{b}})\hat{S}_{\textsc{a}}(\lambda^{\textsc{a}};T)\ket{0}+a_{2}\ket{\mx;T}_{\textsc{a}}\hat{S}_{\textsc{b}}(\lambda^{\textsc{b}})\hat{S}_{\textsc{a}}(-\lambda^{\textsc{a}};T)\ket{0}\right)\\
&+b_{2}\ket{\mx}_{\textsc{b}}\left(a_{1}\ket{\px;T}_{\textsc{a}}\hat{S}_{\textsc{b}}(-\lambda^{\textsc{b}})\hat{S}_{\textsc{a}}(\lambda^{\textsc{a}};T)\ket{0}+a_{2}\ket{\mx;T}_{\textsc{a}}\hat{S}_{\textsc{b}}(-\lambda^{\textsc{b}})\hat{S}_{\textsc{a}}(-\lambda^{\textsc{a}};T)\ket{0}\right).
\end{split}\label{eqb55}
\end{align}
Here we have modified our Gaussian unitary notation, where $\hat{S}_{i}(\lambda^{i};T)$ is the unitary corresponding to detector $i$ (i.e. $F_{i}(\bm{x},\bm{y})$) with interaction strength $\lambda^{i}$ and that has also experienced free-field evolution for time $T$ (i.e. $\hat{U}(T)\hat{S}$). As we can see above, the detectors behave as controls on the quadratic unitaries. The final state, $\ket{\Psi(T^{+})}$, can be viewed as two qubits entangled with Gaussian states.
 
 From \eqref{eqb55} we can construct the corresponding density matrix for the system and trace out the field and Alice's qubit. The resulting reduced density matrix of B can be written as
\begin{align}
\hat{\rho}_{\textsc{b}}&=\begin{pmatrix}
p_{++} & p_{+-}\\
p_{-+} & p_{--}\end{pmatrix},
\end{align}
where
\begin{align}
p_{++}&=\abs{b_{1}}^{2},\\
p_{--}&=\abs{b_{2}}^{2},\\
p_{+-}&=b_{1}\overline{b}_{2} \left( \abs{a_{1}}^{2}\bra{0}\hat{S}_{\textsc{a}}^{\dagger}(\lambda^{\textsc{a}};T)\hat{S}_{\textsc{b}}(2\lambda^{\textsc{b}})\hat{S}_{\textsc{a}}(\lambda^{\textsc{a}};T)\ket{0}
+\abs{a_{2}}^{2}\bra{0}\hat{S}_{\textsc{a}}^{\dagger}(-\lambda^{\textsc{a}};T)\hat{S}_{\textsc{b}}(2\lambda^{\textsc{b}})\hat{S}_{\textsc{a}}(-\lambda^{\textsc{a}};T)\ket{0}\right),\label{eqb59}\\
p_{-+}&=b_{2}\overline{b}_{1} \left( \abs{a_{1}}^{2}\bra{0}\hat{S}_{\textsc{a}}^{\dagger}(\lambda^{\textsc{a}};T)\hat{S}_{\textsc{b}}(-2\lambda^{\textsc{b}})\hat{S}_{\textsc{a}}(\lambda^{\textsc{a}};T)\ket{0}
+\abs{a_{2}}^{2}\bra{0}\hat{S}_{\textsc{a}}^{\dagger}(-\lambda^{\textsc{a}};T)\hat{S}_{\textsc{b}}(-2\lambda^{\textsc{b}})\hat{S}_{\textsc{a}}(-\lambda^{\textsc{a}};T)\ket{0}\right).\label{eqb60}
\end{align}
Note that this matrix is in the x-basis.

\begin{figure}[!t]
\centering
\begin{minipage}{0.485\textwidth}
\centering
\includegraphics[width=\linewidth]{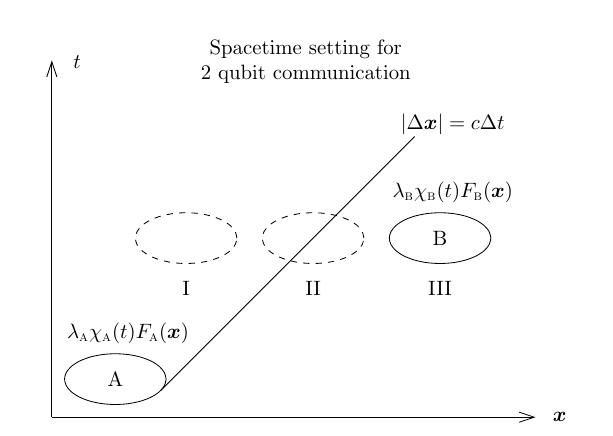}
\caption{\label{fig1}Spacetime setup of the simple 2-qubit communication protocol. Also known as the Fermi problem, the setup involves Alice interacting with the field and then studying how Bob's detector responds at different positions relative to Alice. In a 3+1~D massless QFT, Bob should experience no communication when time-like or space-like separated, only capable of detecting Alice's presence when they share light-like separations. We consider initial conditions where Alice starts in the $\ket{+z}$ state and Bob in the $\ket{-z}$ state.}
\end{minipage}\hfill
\begin{minipage}{0.485\textwidth}
\centering
\includegraphics[width=\linewidth]{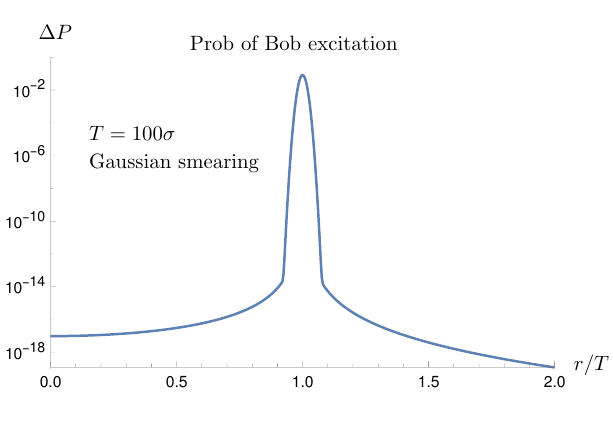}
\caption{\label{fig2}The effect of Alice's interaction on Bob's detector. This plot describes the change in Bob's excitation probability $\ket{-z}\rightarrow\ket{+z}$, due to the presence of Alice's detector (i.e. $P_{\textsc{b}}(\lambda_{\textsc{a}}=1)-P_{\textsc{b}}(\lambda_{\textsc{a}}=0)$), as a function of displacement. Alice and Bob are separated by a distance $r$, and Bob interacts with the field at a time $T=1$. The peak corresponds to Bob's smearing centred on the lightcone emanating from the centre of Alice's smearing. For time-like and space-like separations, this plot does not go to zero as we are using Gaussian smearings that are strictly speaking non-local. However, the log plot reveals the Gaussian suppressed nature of this communication (especially for $r>1.2$). For this calculation, Alice and Bob have Gaussian smearings with $\sigma=10^{-2}$. Also, Alice's qubit is initially excited $\ket{+z}$.} 
\end{minipage}
\end{figure}

In the equations above \eqref{eqb59} and \eqref{eqb60}, the $p_{\pm \mp}$ terms are the off-diagonal, coherence terms of Bob's reduced density matrix. The coherence of Bob's qubit relies on an accurate evaluation of the vacuum expectation value of a product of unitary, quadratic displacement operators, i.e. $\bra{0}\hat{S}_{1}^{\vphantom{\dagger}}\hat{S}_{2}^{\vphantom{\dagger}}\hat{S}_{1}^{\dagger}\ket{0}$ (recall that $a_{i}$ and $b_{i}$ are initial conditions \eqref{eq_ab_54}, \eqref{eq_ab_55}). Exact evaluation of this expectation value requires the Fock representation, along with the global phase information to ensure the correct complex phase to accompany the $\abs{a_{i}}^{2}$ terms. The loss of this phase information would result in a decoherence of the $p_{\pm \mp}$ terms, effectively a qubit-dephasing channel.

%Mathematically \eqref{eqb59} is non-trivial to evaluate as their are vacuum expectation values of three squeezing operators. In principle, the tedious process of finding the correct nullifier $\bm{K}$ matrix for the state $\hat{S}_{\textsc{a}}^{\dagger}(\lambda^{\textsc{a}};T)\hat{S}_{\textsc{b}}(2\lambda^{\textsc{b}})\hat{S}_{\textsc{a}}(\lambda^{\textsc{a}};T)\ket{0}$ can lead to a procedure for evaluating the phase information that is similar to that found in Section~\ref{sec_phi2}. An alternative process that is faster for simple detector smearings (i.e. $F(\bm{x},\bm{y})$) involves exploiting the phase sensitive Fock representations in the

Mathematically, the evaluation of \eqref{eqb59} requires the determination of the nullifier $\bm{K}$ matrix corresponding to a state generated by a sequence of quadratic displacements. This can then be used in the procedure described above in section~\ref{sec_m_4} to evaluate the phase sensitive value of $p_{+-}$. Numerically, adding an additional quadratic displacement increases the computational cost of $\bm{K}$ by requiring an additional evaluation of a matrix exponential and a matrix multiplication. The matrix exponential evaluation can be simplified if the quadratic Gaussian unitary is generated by a simple or known Hamiltonian, e.g. Sec~\ref{sec_phi2}.

 We consider initial conditions where Alice starts in the $\ket{+z}$ state ($a_{1}=a_{2}=1/\sqrt{2}$) and Bob in the $\ket{-z}$ state ($b_{1}=-b_{2}=1/\sqrt{2}$). For the simple case of a Gaussian $F(\bm{x},\bm{y})$,
 \begin{align}
F_{i}(\bm{x},\bm{y})&=\frac{1}{(2\pi)^{3}\sigma^{6}} e^{-\frac{\abs{\bm{x}-\bm{a}_{i}}^{2}}{2\sigma^{2}}-\frac{\abs{\bm{y}-\bm{a}_{i}}^{2}}{2\sigma^{2}}},
 \end{align}
where $\bm{a}_{i}$ is the centre of Alice or Bob's detector. Fig.~\ref{fig2} shows a plot of Alice's contribution to the probability that Bob's qubit becomes excited ($\ket{-z}\rightarrow\ket{+z}$), i.e. $P_{\textsc{b}}(\lambda_{\textsc{a}}=1)-P_{\textsc{b}}(\lambda_{\textsc{a}}=0)$. The plot shows a strong peak at $r=1$, which corresponds to Alice's and Bob's interactions being light-like separated. The tails outside of these peaks are a result of the non-local nature of Gaussian smearing functions, which we justify as exponentially small and therefore can be neglected. There are also numerical errors introduced from the finite precision used by Mathematica in this evaluation, most notably for $r\lesssim 0.8$ where errors of $\sim 10^{-18}$ persist.
 
Broadly speaking, fig.~\ref{fig2} shows that when Bob's interaction is synchronised with the arrival of Alice's signal, the probability of excitation increases significantly. This is a simple protocol used to determine if a modified field theory is still causal, but has never been evaluated with quadratic detectors due to the inability to evaluate the Fock phase term. 

\end{widetext}

\section{Discussion}\label{sec_m_6}
The application of quantum optical techniques in quantum field theory, specifically Gaussian quantum mechanics, has often been a helpful tool for performing non-perturbative calculations. This technique has been practical when considering linearly coupled detectors, exploiting the well-known formula for the inner product of coherent states. The extension of Gaussian QM techniques to quadratically coupled detectors requires an analogous formula describing the inner product between general Gaussian states, a problem that is equivalent to evaluating the phase-sensitive Fock representation of a unitarily generated Gaussian state. Starting from the nullifier equation approach~\cite{K_graphs_squeezed}, our results show how to non-perturbatively determine the exact phase necessary to equate the Fock and unitary representations of a zero-mean Gaussian state. Such a result is necessary when modelling coherent superpositions of Gaussian states, which arise when considering quadratically coupled detectors, i.e. qubit-controlled quadratic unitaries.

As presented in the text above, the phase equating the Fock and unitary representations of a Gaussian state is described by a matrix equation~\eqref{eq40}. For the specific case of a $\phi^{2}$ interaction, this matrix equation simplifies to~\eqref{eq52}. Whilst this result avoids the use of the BCH formula, the evaluation of the phase term requires the calculation of a matrix logarithm. Therefore, instead of recursively using the BCH formula, the problem of evaluating the phase term becomes an eigenvalue problem. 

The result presented above is particularly important for interacting theories that involve superpositions of Gaussian states~\cite{trapped_ion_qubit_squeezing,matsos2024universalquantumgateset,two-mode_squeezed_superpositions}. In relativistic quantum information several protocols meet this criterion if we consider quadratic detectors; these include entanglement harvesting~\cite{a_sachs_2UDW_1}, quantum energy teleportation~\cite{Hotta_2010}, the Fermi problem~\cite{Fermi_RWA}, and the energy density distribution within the QFT post-interaction~\cite{Funai_QET}. In general quantum information, this result is modelling a technique that will become more common as we consider more qubit-controlled quadratic CV couplings~\cite{trapped_ion_qubit_squeezing}, which are expected to arise whilst studying hybrid-QC models~\cite{qubit_beam_split,trapped_ion_universal_QC} and superconducting circuits~\cite{cQED_harvest} in the near future.

Moving further afield, this technique may be useful when modelling gravitational effects of mass superpositions, i.e. spatial cat states. The approximate superposition of quadratic Gaussian states are expected to arise in mass interferometry experiments and also in analogue gravity simulators~\cite{Bei_Lok_Charis_GR_Decoherence,Optomech_GR_decoherence}, e.g. in a superconducting circuits framework~\cite{SineKG_LC_CIRCUIT}.

\section{Conclusion}\label{sec_m_7}
In this manuscript, we have introduced a method for evaluating the exact phase necessary to equate the unitary and Fock state representations of Gaussian states in relativistic, bosonic QFTs. For the specific case of a $\phi^{2}$ interaction, we have demonstrated that evaluation of the phase requires evaluating the trace of a matrix logarithm, a considerably simpler task than using a recursive BCH formula approach.

We also used this new technique to model a simple yet fundamental protocol in relativistic quantum information, namely the Fermi problem. This problem contains mathematics at the heart of all relativistic quantum information protocols and therefore allows the exploration of quadratic RQI protocols, e.g. quantifying the channel capacity of a quadratic communication protocol. This worked example has also demonstrated that two quadratic detector protocols can produce non-divergent results, and this opens the way for quantitative analysis of more elaborate quadratically coupled models.

\section{Acknowledgements}

I would like to thank Ben Baragiola and Nicolas Menicucci for their useful discussions. This work was supported by the Australian Research Council Centre of Excellence for Quantum Computation and Communication Technology (Project No.~CE170100012) and the Australian Research Council Discovery Program (Project No.~DP200102152). 

\widetext

\appendix

\section{Fock representation and Nullifier equations}\label{app_sec_a}
\subsubsection*{Quantum optics regime}
In conventional quantum optics, single-mode squeezed states take the form
\begin{align}
\hat{S}(z)\ket{0}&=e^{-\frac{1}{2}\left(z\hat{b}^{\dagger 2}-z^{*}\hat{b}^{\vphantom{\dagger} 2}\right)}\ket{0},
\end{align}
where the squeeze operator $\hat{S}$ satisfies well known conjugation relations $\hat{S}^{\dagger}(z)\hat{b}\hat{S}(z)=\hat{b}^{\phantom{\dagger}}\text{cosh}(r)-e^{\ii\theta}\hat{b}^{\dagger}\text{sinh}(r)$, where $z=re^{\ii\theta}$. Note that here we use $\hat{b}$ to describe quantum optical modes, i.e. $[\hat{b},\hat{b}^{\dagger}]=1$. This distinguishes them from $\hat{a}_{\bm{k}}$ modes, which are Dirac-delta normalised.

When considering more general quadratic Gaussians (e.g. multimode and including shearing-like terms), the conjugation relations follow a similar general form. Consider a general quadratic Gaussian unitary $\hat{S}(\lambda)$:
\begin{align}
\hat{S}(\lambda)&=\exp\left[-\ii\lambda\sum_{\bm{k},\bm{k}'}\left(A_{\bm{k}\bm{k}'}^{\vphantom{*}}\hat{b}^{\vphantom{\dagger}}_{\bm{k}}\hat{b}^{\vphantom{\dagger}}_{\bm{k}'}+2B_{\bm{k}\bm{k}'}\hat{b}_{\bm{k}}^{\dagger}\hat{b}^{\vphantom{\dagger}}_{\bm{k}'}+A^{*}_{\bm{k}\bm{k}'}\hat{b}_{\bm{k}}^{\dagger}\hat{b}_{\bm{k}'}^{\dagger}\right)\right],
\end{align}
where the sums over $\bm{k}$ and $\bm{k}'$ enumerate the $N$ optical modes relevant to the problem. This leads to conjugation relation (or similarity transformations) of the form
\begin{align}
\hat{S}(-\lambda)\left(\bm{G}_{1}^{\vphantom{-1}}\hat{\bm{b}}+\bm{G}_{2}^{\vphantom{-1}}\hat{\bm{b}}^{\dagger}\right)\hat{S}(\lambda)=&\begin{pmatrix}
\bm{G}_{1}^{\vphantom{-1}} & \bm{G}_{2}^{\vphantom{-1}}
\end{pmatrix}
\text{exp}\left[2\ii\lambda
\begin{pmatrix}
-\bm{B} & -\bm{A}^{*} \\
\bm{A} & \bm{B}^{*}
\end{pmatrix}\right]
\begin{pmatrix}
\hat{\bm{b}}\\
\hat{\bm{b}}^{\dagger}
\end{pmatrix},\label{eq_a_a3}
\end{align}
where $\hat{\bm{b}}$ is the vector of annihilation operators and $\bm{G}_{1}^{\vphantom{-1}},\bm{G}_{2}^{\vphantom{-1}}\in\text{Mat}_{N\times N}(\mathbb{C})$. This relation can be derived using the \textit{Lie algebraic similarity transformation method} described in~\cite{puri2001mathematical}. This conjugation property is extremely useful and, ordinarily, sufficient for dealing with Gaussian operations in quantum optics. 

A particularly useful application of \eqref{eq_a_a3} is to consider the choice
\begin{align}
\begin{pmatrix}
\bm{G}_{1}^{\vphantom{-1}} & \bm{G}_{2}^{\vphantom{-1}}
\end{pmatrix}&=
\begin{pmatrix}
\bm{E} & \bm{0}
\end{pmatrix}
\text{exp}\left[-2\ii\lambda
\begin{pmatrix}
-\bm{B} & -\bm{A}^{*} \\
\bm{A} & \bm{B}^{*}
\end{pmatrix}\right],
\end{align}
where $\bm{E}$ is an invertible $N\times N$ matrix. Then
\begin{align}
\hat{S}(-\lambda)\left(\bm{G}_{1}^{\vphantom{-1}}\hat{\bm{b}}+\bm{G}_{2}^{\vphantom{-1}}\hat{\bm{b}}^{\dagger}\right)\hat{S}(\lambda)&=\bm{E}\hat{\bm{b}}.
\end{align}
If we consider this result in tandem with a quadratic Gaussian state generated by $\hat{S}(\lambda)$, we can see that:
\begin{align}
\left(\bm{G}_{1}^{\vphantom{-1}}\hat{\bm{b}}+\bm{G}_{2}^{\vphantom{-1}}\hat{\bm{b}}^{\dagger}\right)\hat{S}(\lambda)\ket{0}&=\hat{S}(\lambda)\hat{S}(-\lambda)\left(\bm{G}_{1}^{\vphantom{-1}}\hat{\bm{b}}+\bm{G}_{2}^{\vphantom{-1}}\hat{\bm{b}}^{\dagger}\right)\hat{S}(\lambda)\ket{0},\\
&=\hat{S}(\lambda)\bm{E}\hat{\bm{b}}\ket{0}=0.
\end{align}
By using the conjugation properties of the quadratic Gaussian, we have found the $N$ (linearly independent) linear operators $\bm{G}_{1}^{\vphantom{-1}}\hat{\bm{b}}+\bm{G}_{2}^{\vphantom{-1}}\hat{\bm{b}}^{\dagger}$ that annihilate the state $\hat{S}(\lambda)\ket{0}$. Thus, the equation above is known as the nullifier equation~\cite{K_graphs_squeezed}, and the operator $\bm{G}_{1}^{\vphantom{-1}}\hat{\bm{b}}+\bm{G}_{2}^{\vphantom{-1}}\hat{\bm{b}}^{\dagger}$ is known as the nullifier of the Gaussian state.

The nullifier equation plays a central role in the main text as it can be rearranged into a more revealing form:
\begin{align}
\left(\hat{\bm{b}}+\bm{G}_{1}^{-1}\bm{G}_{2}^{\vphantom{-1}}\hat{\bm{b}}^{\dagger}\right)\hat{S}(\lambda)\ket{0}&=0,\\
\left(\hat{\bm{b}}-\bm{K}\hat{\bm{b}}^{\dagger}\right)\hat{S}(\lambda)\ket{0}&=0,
\end{align}
where $\bm{K}=-\bm{G}_{1}^{-1}\bm{G}_{2}^{\vphantom{-1}}$. Note that $\bm{K}$ is independent of the choice of $\bm{E}\in \text{GL}_{N}(\mathbb{C})$. We can now interpret the state $\hat{S}(\lambda)\ket{0}$ as an element of the kernel of the nullifier $(\hat{\bm{b}}-\bm{K}\hat{\bm{b}}^{\dagger})$, and use this fact to describe the state $\hat{S}(\lambda)\ket{0}$~\cite{K_graphs_squeezed}. By using a Fock basis expansion, one can show that the kernel of the nullifier is a 1D vector space, whose elements are given by:
\begin{align}
\left(\hat{\bm{b}}-\bm{K}\hat{\bm{b}}^{\dagger}\right)\mathcal{N}e^{\frac{\hat{\bm{b}}^{\textsc{h}}\bm{K}\hat{\bm{b}}^{\dagger}}{2}}\ket{0}&=0,
\end{align}
where $\mathcal{N}\in\mathbb{C}$ and $\hat{\bm{b}}^{\textsc{h}}$ is a row vector consisting of creation operators. 

Since $\hat{S}(\lambda)\ket{0}$ is an element of the kernel of $(\hat{\bm{b}}-\bm{K}\hat{\bm{b}}^{\dagger})$, and this nullifier has a 1D kernel space we can therefore write the Fock representation of the Gaussian state as:
\begin{align}
\hat{S}(\lambda)\ket{0}&=\mathcal{N}e^{\frac{\hat{\bm{b}}^{\textsc{h}}\bm{K}\hat{\bm{b}}^{\dagger}}{2}}\ket{0},
\end{align}
where $\mathcal{N}$ is a complex scalar. The absolute value of $\mathcal{N}$ can be calculated easily by recalling that $\hat{S}(\lambda)\ket{0}$ is a unit length vector, however the phase of $\mathcal{N}$ is a more challenging problem and is the result presented in the main text above.
\subsubsection*{Quantum field theory regime}
When considering QFT modes the ladder operators are Dirac-delta normalised, i.e. $[\hat{a}_{\bm{\kappa}}^{\vphantom{\dagger}},\hat{a}_{\bm{\lambda}}^{\dagger}]=\delta_{\bm{\kappa\lambda}}/\bm{\Delta k}$. This introduced minor alterations in the results presented above, mostly via appropriate introductions of $\bm{\Delta k}$ factors:
\begin{align}
\hat{S}_{\textsc{qft}}(\lambda)&=\exp\left[-\ii\lambda\bm{\Delta k}^{2}\sum_{\bm{k},\bm{k}'}\left(A_{\bm{k}\bm{k}'}^{\vphantom{*}}\hat{a}^{\vphantom{\dagger}}_{\bm{k}}\hat{a}^{\vphantom{\dagger}}_{\bm{k}'}+2B_{\bm{k}\bm{k}'}\hat{a}_{\bm{k}}^{\dagger}\hat{a}^{\vphantom{\dagger}}_{\bm{k}'}+A^{*}_{\bm{k}\bm{k}'}\hat{a}_{\bm{k}}^{\dagger}\hat{a}_{\bm{k}'}^{\dagger}\right)\right],\\
\hat{S}_{\textsc{qft}}(-\lambda)\left(\bm{G}_{1}^{\vphantom{-1}}\hat{\bm{a}}+\bm{G}_{2}^{\vphantom{-1}}\hat{\bm{a}}^{\dagger}\right)\hat{S}_{\textsc{qft}}(\lambda)&=\begin{pmatrix}
\bm{G}_{1}^{\vphantom{-1}} & \bm{G}_{2}^{\vphantom{-1}}
\end{pmatrix}
\text{exp}\left[2\ii\lambda\bm{\Delta k}
\begin{pmatrix}
-\bm{B} & -\bm{A}^{*} \\
\bm{A} & \bm{B}^{*}
\end{pmatrix}\right]
\begin{pmatrix}
\hat{\bm{a}}\\
\hat{\bm{a}}^{\dagger}
\end{pmatrix}.
\end{align}
The useful choice of $\bm{G}_{1}^{\vphantom{-1}}$ and $\bm{G}_{2}^{\vphantom{-1}}$ becomes
\begin{align}
\begin{pmatrix}
\bm{G}_{1}^{\vphantom{-1}} & \bm{G}_{2}^{\vphantom{-1}}
\end{pmatrix}&=
\begin{pmatrix}
\bm{E} & \bm{0}
\end{pmatrix}
\text{exp}\left[-2\ii\lambda\bm{\Delta k}
\begin{pmatrix}
-\bm{B} & -\bm{A}^{*} \\
\bm{A} & \bm{B}^{*}
\end{pmatrix}\right],
\end{align}
such that
\begin{align}
\left(\hat{\bm{a}}+\bm{G}_{1}^{-1}\bm{G}_{2}^{\vphantom{-1}}\hat{\bm{a}}^{\dagger}\right)\hat{S}_{\textsc{qft}}(\lambda)\ket{0}&=0,\\
\left(\hat{\bm{a}}-\bm{\Delta k}\bm{K}\hat{\bm{a}}^{\dagger}\right)\hat{S}_{\textsc{qft}}(\lambda)\ket{0}&=0,\\
\left(\hat{\bm{a}}-\bm{\Delta k}\bm{K}\hat{\bm{a}}^{\dagger}\right)\mathcal{N}e^{\bm{\Delta k}^{2}\frac{\hat{\bm{a}}^{\textsc{h}}\bm{K}\hat{\bm{a}}^{\dagger}}{2}}\ket{0}&=0,
\end{align}
Therefore:
\begin{align}
\hat{S}_{\textsc{qft}}(\lambda)\ket{0}&=\mathcal{N}e^{\bm{\Delta k}^{2}\frac{\hat{\bm{a}}^{\textsc{h}}\bm{K}\hat{\bm{a}}^{\dagger}}{2}}\ket{0}.
\end{align}
This result carries through in the $\bm{\Delta k}\rightarrow\bm{0}$ limit and therefore faithfully describes a quadratic Gaussian state in a cavity-less QFT.

\section{Fields in cavities}\label{sec_a_b}
\subsubsection*{Mode decomposition}
The results presented in this manuscript require a momentum space with discrete or countable dimensions. Instead of an arbitrary discretisation of the momentum space, we exploit previous work~\cite{Funai_thesis} to use a physically motivated and convergent discretisation, i.e. considering the quantum field inside a rectangular cavity. We consider a cavity with three spatial dimensions and size $x_{i}\in[0,L_{i}]$, with Dirichlet boundary conditions. 

Given this setup, the quantum field can be decomposed into Fourier modes of the form
\begin{align}
\hat{\phi}(t,\bm{x})=&\sum_{\bm{\kappa}}
\bm{\Delta k}
\sqrt{\frac{4}{\omega\pi^{3}}}\left(\hat{a}_{\bm{\kappa}}^{\vphantom{\dagger}}e^{-\ii\omega t}+\hat{a}_{\bm{\kappa}}^{\dagger}e^{\ii\omega t}\right)\sin\kappa_{1}x_{1}\sin\kappa_{2}x_{2}\sin\kappa_{3}x_{3} \tag{$\phi$ plane wave},\\
\hat{\pi}(t,\bm{x})=&-\ii\sum_{\bm{\kappa}}
\bm{\Delta k}
\sqrt{\frac{4\omega}{\pi^{3}}}\left(\hat{a}_{\kappa}^{\vphantom{\dagger}}e^{-\ii\omega t}-\hat{a}_{\kappa}^{\dagger}e^{\ii\omega t}\right)\sin\kappa_{1}x_{1}\sin\kappa_{2}x_{2}\sin\kappa_{3}x_{3} \tag{$\pi$ plane wave},
\end{align}
where $\bm{\Delta k}=\pi/V$ ($V$ is the cavity volume) and $\kappa_{i}=n_{i}\pi/L_{i}$, $n_{i}\in\mathbb{N}$. Since we are using a massless field, we have $\omega=\abs{\bm{\kappa}}$. These equations can be inverted to find the ladder operators:
\begin{align}
\hat{a}_{\bm{\kappa}}^{\vphantom{\dagger}}=&\int\limits_{V}\dd^{3}\bm{x}\,\frac{2}{\pi^{3/2}}\left(\sqrt{\omega}\hat{\phi}(t,\bm{x})+\frac{\ii}{\sqrt{\omega}}\hat{\pi}(t,\bm{x})\right)\sin\kappa_{1}x_{1}\sin\kappa_{2}x_{2}\sin\kappa_{3}x_{3}e^{\ii\omega t},\tag{creation operator}\\
\hat{a}_{\bm{\kappa}}^{\dagger}=&\int\limits_{V}\dd^{3}\bm{x}\,\frac{2}{\pi^{3/2}}\left(\sqrt{\omega}\hat{\phi}(t,\bm{x})-\frac{\ii}{\sqrt{\omega}}\hat{\pi}(t,\bm{x})\right)\sin\kappa_{1}x_{1}\sin\kappa_{2}x_{2}\sin\kappa_{3}x_{3}e^{-\ii\omega t} \tag{annihilation operator},
\end{align}
where the integral domain is the cavity interior $x_{i}\in[0,L_{i}]$. As usual, the field operators obey Dirac-normalised canonical commutation relations, hence
\begin{align}
\left[\hat{\phi}(t,\bm{x}),\hat{\pi}(t,\bm{y})\right]&=\ii\delta(\bm{x}-\bm{y}),\\
\left[\hat{a}^{\vphantom{\dagger}}_{\bm{\kappa}},\hat{a}^{\dagger}_{\bm{\lambda}}\right]&=\frac{\delta_{\bm{\kappa \lambda}}}{\bm{\Delta k}},
\end{align}
where $\delta_{\bm{\kappa \lambda}}$ is a Kronecker delta. The mode decomposition described above is used to evaluate equations \eqref{eq_z_15}-\eqref{eq_z_17}. 
\subsubsection*{Cavities vs free space fields}
This approach of physically discretising momentum has some advantages over general mathematical discretisation. An analysis of the 2-point correlator (Wightman function) in a cavity field reveals that the effects of the cavity walls decay as $1/L^{2}$ for `experiments' far from the wall, i.e. $\bm{x}\approx \bm{L}/2$. Given the omnipresence of the Wightman function in QFT (free and interacting), this result can be used to bound the error introduced by the presence of the cavity walls. As a consequence, the physics occurring in a cavity field (far from the walls) converges to that of a non-cavity field polynomially. We have exploited this fact in the main text to simplify certain steps in the calculation of the main result. For more details regarding this convergence, see chapter 4 of~\cite{Funai_thesis}.

\twocolumngrid

\bibliography{ref_1}

\end{document}